# Anomalous Subsurface Vacancy Stabilization Dictated by Geometry–Electronic Decoupling on Metal Surfaces

*Yiming Tan*[1], *Pai Li*,[*1]

[1]*State Key Laboratory of Materials for Integrated Circuits Shanghai Institute of Microsystem and Information Technology Chinese Academy of Sciences Shanghai 200050, China*

***Corresponding Author**

Email: lipai@mail.sim.ac.cn

## ABSTRACT

Vacancy formation energetics fundamentally govern the structural integrity and catalytic behavior of metal surfaces. Contrary to conventional coordination-dependent broken-bond models, we identify an anomalous thermodynamic inversion on close-packed surfaces across Ir, Pt, Au with face-centered cubic (FCC) lattice and Be, Zn, Cd with hexagonal close-packed (HCP) lattice, where subsurface vacancies are intrinsically more stable than surface ones. Using high-throughput DFT calculations and machine learning force fields, we demonstrate that the physical origins of this anomaly are fundamentally decoupled between the two crystal systems. For the FCC trio, pronounced surface relaxation and a profound real-space electronic localization induce directional, covalent-like intralayer bonding, materializing a "geometry–electronic decoupling" mechanism. Crucially, this unconventional thermodynamic hierarchy enables a dynamic "self-healing" mechanism on Pt(111) that preserves an intact topmost layer and prevents catalytic degradation during hydrogen evolution and oxygen reduction reactions. It also successfully decoding the critical defect threshold (~8%) for lifting the Au(100) surface reconstruction. Our work challenges classical scalar defect models and provides a fresh paradigm for engineering catalyst surface integrity.

**KEYWORDS**

Subsurface vacancies; Transition metal surfaces; Density functional theory; Self-healing mechanism; Heterogeneous catalysis.

## INTRODUCTION

Transition metals are indispensable cornerstones in heterogeneous catalysis[1,2], low-dimensional materials growth[3,4], and surface self-assembly[5,6], serving as primary active components in a wide range of energy-relevant chemical transformations[7]. The catalytic performance and long-term durability of these metals are intrinsically governed by the atomic-scale topography of their surfaces, which is highly sensitive to the presence of structural defects[8,9]. Among various defects, atomic vacancies play a central role in modulating the local electronic structure, driving surface stress relaxation, and determining the dynamic reconstruction pathways under realistic operating conditions. Consequently, achieving a quantitative, predictive grasp of vacancy formation energetics is an essential prerequisite for the rational design of high-performance catalysts[10].

Conventionally, vacancy stability on transition metal surfaces is interpreted through the lens of the coordination-dependent broken-bond model. Within this classical framework, the energetic cost of defect formation is scaled almost linearly with the number of chemical bonds severed upon atom removal[11,12]. Because surface atoms possess fewer neighbors than subsurface or bulk atoms, they are universally expected to exhibit significantly lower vacancy formation energies. This simple yet powerful scalar paradigm has been validated across a broad spectrum of metallic systems and has long served as an unquestioned cornerstone for interpreting defect thermodynamics, surface diffusion, and interfacial chemistry[13,14].

However, recent experimental observations have begun to shake this foundational assumption. Crucially, scanning tunneling microscopy (STM) and structural characterizations have unexpectedly revealed the stable existence of buried subsurface vacancies beneath the densified outermost layers of reconstructed Au systems[15,16]. This inverted energetic hierarchy—where creating a defect deeper in the lattice is energetically cheaper than doing so at the surface—directly violates the conventional coordination-based paradigm. It strongly implies that the vacancy stability anomaly is an intrinsic phenomenon governed by sophisticated electronic effects that escape simple bond-counting arguments. While this anomaly is highly intriguing, a unified, generalized understanding of its elemental distribution across the periodic table and its underlying microscopic driving force has hitherto remained elusive.

In this work, we systematically map the landscape of vacancy formation energies across a comprehensive suite of FCC and HCP metals using high-throughput DFT calculations. We discover that this thermodynamic inversion is not an isolated anomaly but a widespread property that splits into two unique branches: the FCC noble metals (Ir, Pt, Au) and the main-group/transition HCP metals (Be, Zn, Cd). By exploiting the atom-resolved energy decomposition of advanced neural network-based machine learning potentials alongside electronic topology analyses, we demonstrate that classical empirical methods fail entirely because they neglect directional bonding interactions.

We show that for the FCC anomalies (exemplified by Pt), a profound real-space electronic localization yields sharp, covalent-like intralayer bonding channels along the in-plane atomic axes, creating an exceptionally robust surface network that penalizes out-of-plane disruption. This phenomenon beautifully materializes a "geometry–electronic decoupling," where structurally similar local environments exhibit radically different electronic behaviors. Furthermore, we bridge this fundamental thermodynamic discovery with realistic applications by evaluating the Gibbs free energy profiles for the hydrogen evolution reaction (HER) and oxygen reduction reaction (ORR) on Pt(111), unveiling a dynamic "self-healing" mechanism that maintains a flat, active catalytic layer. Finally, we utilize this framework to successfully decode the energetic threshold and critical defect concentration (~8%) required to lift the mysterious Au(100) surface reconstruction under particle bombardment, completing a comprehensive closed-loop framework from quantum mechanical origin to macroscopic catalysis and phase transition.

# RESULTS AND DISCUSSION

## 1. Stability of surface and subsurface vacancies in typical FCC and HCP metals

To systematically investigate vacancy formation energetics across different crystal structures, we focus on close-packed surfaces of transition metals, specifically the (111) surfaces for typical FCC metals and the (0001) surfaces for HCP metals. As illustrated in **Figure 1**, for the majority of metals, the formation energy of a subsurface vacancy is higher than that of a surface vacancy ($\Delta E_f = E_f^{sub} - E_f^{surf} > 0$), which is consistent with the classical bond-breaking model. However, we identify three FCC metals (Pt and Au, with a weaker tendency for Ir) and three HCP metals (Be, Zn and Cd) that exhibit an abnormal stability hierarchy with $\Delta E_f < 0$. Notably, the identified FCC metals are all late transition metals in the 5*d* series. For the HCP metals, the electronic configurations vary: Be possesses filled 1$s$ and 2$s$ subshells, whereas Zn and Cd feature filled *d* and *s* shells. To ensure that these observations are not artifacts of computational methodology, we performed a robustness analysis using various exchange-correlation functionals, including PBE[17], PBEsol[18], SCAN[19], and HSE[20,21], both with and without D3 dispersion corrections[22,23]. For Pt, the negative value of $\Delta E_f$ remains consistent across all functionals, confirming that the preference for subsurface vacancies is an intrinsic physical property.

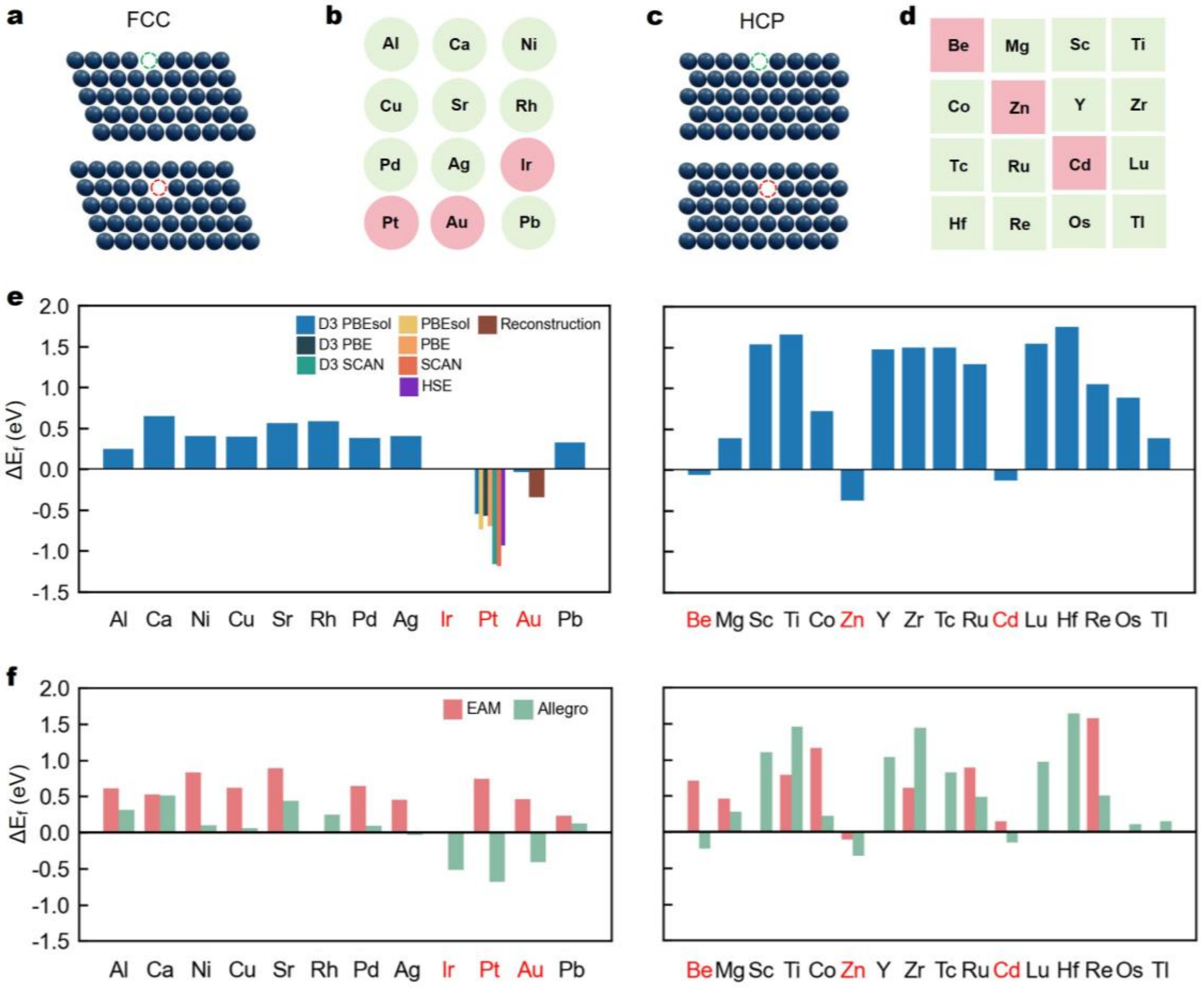

**Figure 1.** Structural models and vacancy formation energetics on close-packed surfaces of FCC and HCP metals. (a, c) Schematic illustrations of a surface vacancy (green dashed circle) and a subsurface vacancy (red dashed circle) on the FCC (111) and HCP (0001) surfaces, respectively. (b, d) High-throughput screening maps showing the distribution of normal (green) and anomalous (red) metals. (e) Vacancy formation energy differences ($\Delta E_f = E_f^{sub} - E_f^{surf}$) for various FCC and HCP metals calculated using different exchange-correlation functionals. For Pt, the negative $\Delta E_f$ is consistently reproduced across multiple functionals and surface conditions (including reconstruction). For Au(111), $\Delta E_f$ of the 22×√3 reconstructed surface is also provided. (f) Comparison of $\Delta E_f$ predicted by EAM and Allegro potentials.

We note that this anomalous phenomenon is deeply tied to the structural density of close-packed surface lattices. On the pristine, unreconstructed (100) surfaces of FCC metals (**Figure S1**), a surface vacancy universally remains thermodynamically more stable than a subsurface one. This compliance with the classical broken-bond model arises because the square lattice of the open (100) surface lacks the strong, directional, covalent-like intralayer interactions characteristic of the pristine FCC (111) or HCP (0001) facets. Remarkably, however, this energetic hierarchy flips drastically below the zero baseline when accounting for the experimental surface phase transition of gold—specifically, the densified, quasi-hexagonal reconstruction of the Au(100) surface (about -0.35 eV, **Figure S1** and also reported in literature[16]). This striking inversion firmly demonstrates that shifting the surface symmetry from an open square to a close-packed hexagonal topology acts as a decisive structural trigger that immediately activates the electronic reconstruction and expels defects into the subsurface.

**2. Origin of the anomalous stability**

To systematically assess the thermodynamic origins of this anomaly and benchmark the predictive fidelity of different potential models, we contrasted our high-throughput DFT calculations against both the classical Embedded Atom Method (EAM)[24,25] and the modern, state-of-the-art machine learning force field (MLFF) Allegro[26,27] (**Figure 1**f). The classical EAM formulation partitions the total energy into a directionless electron density embedding term and a short-range pairwise repulsive interaction, effectively treating metallic bonding as the embedding of ions into an isotropic electron sea. By omitting directional, covalent-like pair interactions, EAM notoriously fails to describe complex, symmetry-breaking surface phase transitions, such as the herringbone reconstruction of Au(111). Paralleling this physical limitation, EAM completely fails to reproduce the subsurface vacancy preference across the anomalous FCC trio (Ir, Pt, and Au) (**Figure 1**f, left panel), which strongly implies that these specific truncated surfaces harbor a pronounced, non-isotropic directional pair-bonding character that purely coordination-dependent scalar models cannot capture.

Intriguingly, this methodology-dependent inconsistency entirely vanishes for the HCP anomalous systems. EAM successfully reproduces the lower, near-negative $\Delta E_f$ for HCP Zn and Cd. This striking contrast indicates that the abnormal subsurface vacancy stability in HCP metals does not stem from a departure from typical metallic bonding, nor from the ultra-strong, directional intralayer surface interactions that uniquely characterize Pt and Au. Instead, the advanced multi-body neural network architecture of Allegro exhibits superior predictive performance across both crystal systems by seamlessly capturing complex, high-dimensional potential energy surfaces. This benchmark successfully decouples the distinct physical origins driving these instabilities: for the FCC noble metals, the subsurface vacancy preference is intrinsically dictated by a real-space electronic reconstruction and covalent-like localization, whereas for the HCP systems, the interplay of filled-shell electron configurations and local structural geometry yields a fundamentally different thermodynamic landscape.

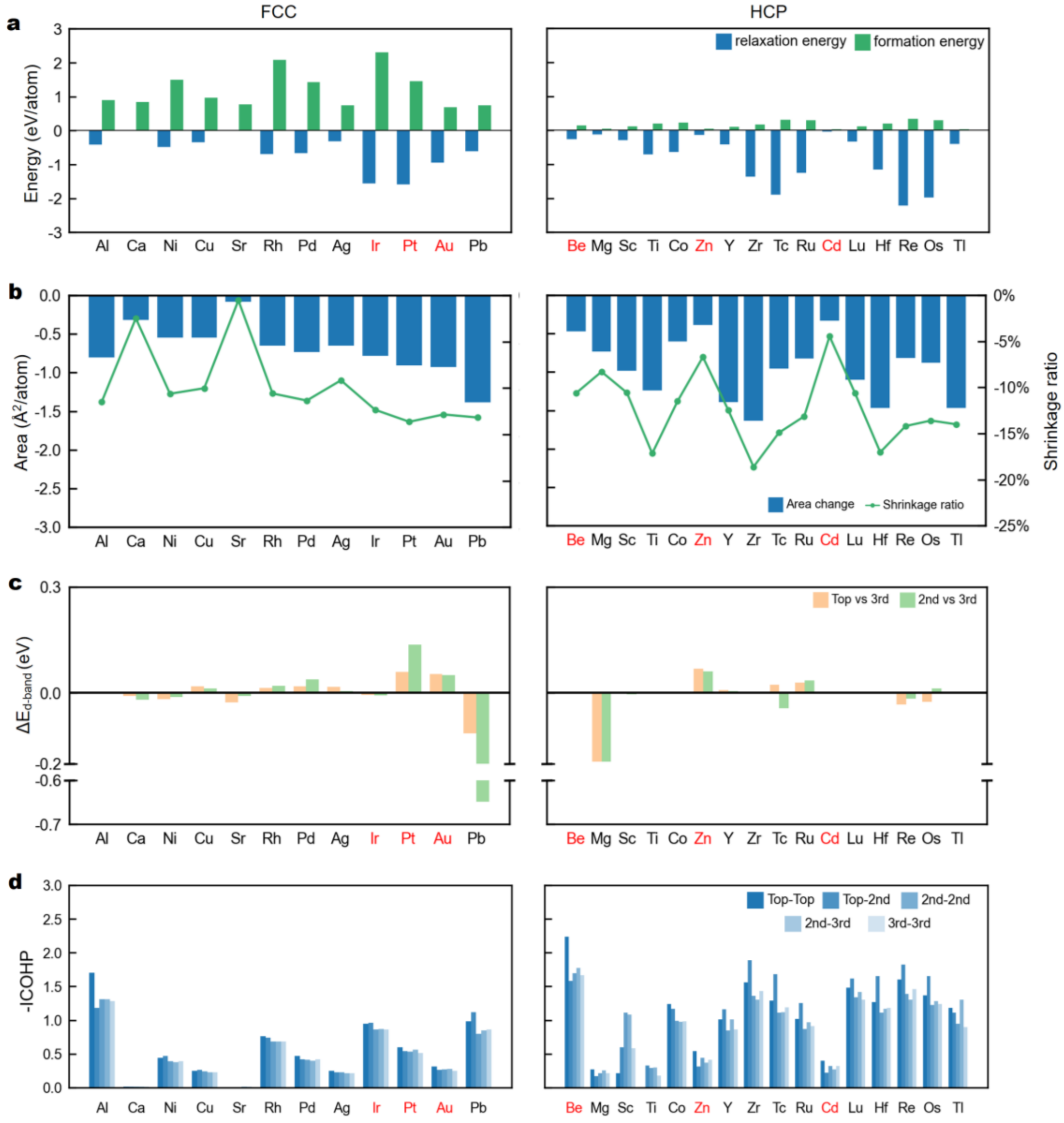

**Figure 2.** Quantitative thermodynamic, structural, and electronic descriptors for close-packed planes of FCC and HCP metals. (a) Extracted structural relaxation energies (blue) and free-standing formation energies (green) for simulated isolated atomic monolayers. (b) Real-space area change (blue bars, left axis) and lateral shrinkage ratio (green connected dots, right axis) of the close-packed monolayers upon structural relaxation. (c) Sub-resolved shifts in the *d*-band center ($\Delta E_{d\text{-}band}$) for the topmost surface (orange) and second subsurface (green) layers relative to the bulk-like inner layer. (d) Layer-dependent -ICOHP demonstrating a universal, monotonic bonding strength evolution from the bulk environment toward the truncated surface across all studied crystalline systems.

Because the neural network-based Allegro potential accurately captures the anomalous thermodynamic trend, we exploited its atom-resolved energy decomposition capability to map the individual atomic energies of surface atoms adjacent to a subsurface vacancy (**Figure S2, S3**). For conventional FCC metals (e.g., Al, Ni, and Rh), the introduction of a subsurface vacancy expectedly elevates the energy of neighboring surface atoms due to local coordination loss, appearing as prominent, high-energy bright localized regions around the defect projection site. However, the three anomalous FCC metals (Ir, Pt, and Au) present striking exceptions: the local energies of their adjacent surface atoms unexpectedly undergo a spontaneous reduction upon subsurface vacancy formation, materializing as suppressed, lower-energy dark regions. This localized thermodynamic response implies that an inward reduction in the out-of-plane coordinates of these specific surface atoms acts to heavily stabilize the outermost layer in Ir, Pt, and Au—a localized self-strengthening effect that is completely absent across the entire HCP series, including Be, Zn, and Cd.

This macroscopic-microscopic correlation allows us to conceptualize the pristine (111) surfaces of these anomalous FCC metals as a *quasi-heterostructure* composed of a strained, close-packed monolayer supported on its bulk substrate, wherein vacancies preferentially segregate to the bulk-like side of the interface rather than the highly robust monolayer. It also illustrates the hexagonal reconstructions of Pt(001) and Au(001) surfaces[28]. Guided by this heterostructure paradigm, we simulated isolated, free-standing close-packed monolayers of these crystalline systems (**Figure 2**a,b). Upon structural relaxation, the Ir, Pt, and Au monolayers exhibit the most dramatic lateral lattice shrinkage, forming a prominent valley in the shrinkage ratio profile (dropping up to 15% for Pt) and yielding a massive structural relaxation energy release exceeding -1.0 eV/atom. While the post-transition metal Pb also shows substantial monolayer contraction, its corresponding energetic gain remains remarkably smaller. In stark contrast, the HCP trio (Be, Zn, and Cd) exhibits minimal geometric shrinkage and negligible relaxation energy lowering, further confirming that the driving forces in the HCP and FCC anomalous sets are fundamentally decoupled.

The underlying electronic distinction is also manifested in the layer-resolved *d*-band center shifts relative to the inner bulk layer ($\Delta E_{d\text{-band}}$). As illustrated in **Figure 2**c, Pt and Au display the most pronounced positive *d*-band center upshifts ($\Delta E_{d\text{-band}} > 0$) driven by strong lateral *d*-orbital compression—a trend widely observed when severe surface tensile stress couples with local electronic alterations[29,30]. In contrast, this *d*-band mediated logic completely breaks down for the HCP systems: Be possesses no *d*-electrons, while the deeply buried, filled d-shells of Zn and Cd remain chemically inert, showing negligible shifts (**Figure 2c**, right panel).

To visually and quantitatively resolve this distinct bonding landscape, we contrast these scalar bonding metrics with real-space topological analyses. Concurrently, Crystal Orbital Hamilton Population (COHP) analysis[31] shows that the integrated COHP (-ICOHP) values and their depth-dependent profiles for Pt exhibit no qualitative deviation from conventional FCC metals (such as Cu and Ni), displaying a universal, monotonic enhance from the bulk toward the surface (**Figure 2**d). Across all pristine FCC slab models, the total bonding strength follows an identical step-like attenuation as it transitions from bulk-like layers to the truncated surface. This paradox strongly demonstrates that scalar, spatially averaged bonding metrics are insufficient to capture the anisotropic electronic reconstruction governing the Pt surface anomaly[32]. Instead, the true driving force lies in the highly directional spatial redistribution of electrons rather than the total scalar bond energy.

This contrast is prominently materialized in the Electron Localization Function (ELF) contour maps and their corresponding 1D linear profiles (**Figure 3**). In their perfect structures, the valence electrons of conventional metals (Ag and Rh) exhibit a nearly isotropic, quasi-spherical ELF distribution centered around the atomic nuclei, resulting in symmetric, unremarkable 1D profiles along the interatomic paths. In sharp contrast, the ELF values of the perfect Pt surface layer peak precisely at the bond centers, tracking a high electron density in directional bonding orbitals. While this bond-centered localization relaxes slightly in the subsurface, it remains significantly stronger than that in Ag, Rh, and Au. Cross-sectional views of the ELF profiles (**Figure S4**) further corroborate this anisotropy: while the interlayer bonding in Pt retains a delocalized, metallic character, the intralayer bonding within the surface layer exhibits a striking electron localization along interatomic axes, confirming the emergence of highly directional, covalent-like interactions. The conclusions derived from the ELF analysis cannot be obtained from either the overall charge distribution maps of different elements or their corresponding 1D profiles (**Figure S5**). The comparison between these two analytical approaches further demonstrates that a holistic analysis fails to capture the essential origin of the anomaly in Pt.

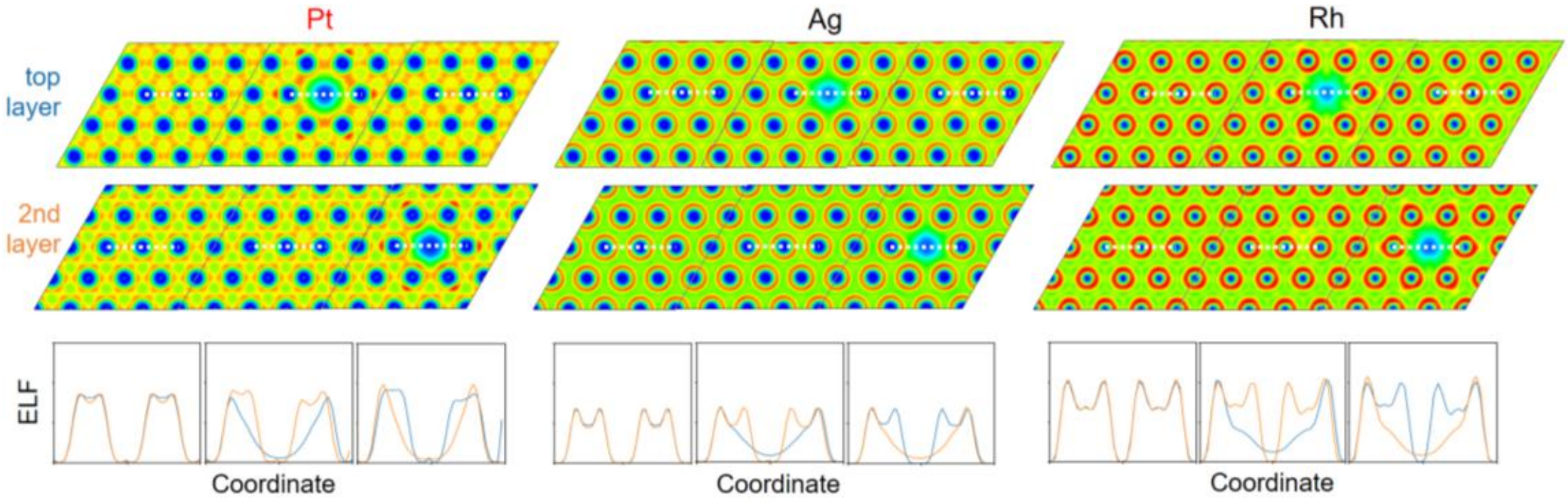

**Figure 3.** Real-space ELF analyses of perfect and defective FCC surfaces. Two-dimensional ELF contour maps are displayed for the top layer (upper row) and 2nd layer (middle row) of Pt, Ag, and Rh. Within each metal panel, from left to right, the subplots illustrate the perfect lattice, the system with a top-layer vacancy, and the system with a 2nd-layer (subsurface) vacancy, respectively. The white dashed lines indicate the linear pathways for the corresponding 1D ELF profiles shown in the bottom row. For Pt, the introduction of a vacancy induces a dramatic, anisotropic electronic reconstruction, where electrons preferentially localize within the sharp bonding channels of the nearest-neighbor hexagonal ring, a feature contrastingly absent in conventional metals (Ag and Rh).

Crucially, this electronic divergence becomes even more pronounced upon defect creation. A comparative analysis of the 2D ELF contour maps and 1D line scans in **Figure 3** reveals that the introduction of a vacancy—whether at the surface or subsurface—triggers a massive, directional electronic reconstruction in Pt that is absent in conventional metals. In Ag and Rh, the local electronic structure is largely insensitive to vacancies, and their 1D ELF paths through the defect regions show a characteristic collapse toward a delocalized background. Conversely, the vacancy-induced excess electrons in Pt preferentially accumulate within the distinct bonding channels of the hexagonal ring formed by the six nearest neighbors surrounding the defect site, as clearly reflected by the sharp, localized peak features in its 1D ELF profiles (**Figure 3**, bottom left panels).

This selective charge enrichment significantly strengthens the lateral cohesion within the close-packed hexagonal network. Such an extraordinary tendency toward robust in-plane bonding renders the Pt surface layer exceptionally rigid. As a consequence, breaking bonds associated with a surface vacancy (which destabilizes this reinforced surface layer) becomes energetically more costly than creating a subsurface vacancy, directly dictating the anomalous thermodynamic inversion ($\Delta E_f < 0$).

Taken together, these results demonstrate that the anomalous vacancy formation energy trends observed in FCC and HCP systems arise from two fundamentally distinct physical mechanisms. The anomaly in HCP systems (Be, Zn, and Cd) is likely governed by non-directional *sp*-hybridization coupled with geometric constraints inherent to hexagonal packing, showing minimal structural or electronic sensitivity to surface creation. Conversely, the FCC anomalies—exemplified by the extreme case of Pt—are driven by a profound, d-electron-mediated electronic reconstruction. This phenomenon beautifully exemplifies a "geometry–electronic decoupling," where the dramatic real-space electron localization breaks away from simple coordination-dependent physics. By centralizing excess electrons into local intralayer bonding channels, Pt creates a self-strengthening surface lattice that penalizes subsurface disruption, ultimately underpinning its anomalous vacancy stability. Given that HCP Be, Zn, and Cd are either highly reactive or toxic, we restrict our primary focus to the FCC systems, particularly Pt and Au, due to their paramount catalytic importance and superior stability.

### 3. Self-healing mechanism on the Pt(111) surface and its catalytic implications

A direct consequence of this thermodynamic inversion—where subsurface vacancies are preferred over surface ones—is the intrinsic "self-healing" capacity of the Pt(111) surface. While Pt is a premier catalyst for both the HER and ORR, surface defects typically act as poisoning centers due to their overly strong coordination with reaction intermediates. Under external perturbations that strip away a surface Pt atom, a neighboring subsurface atom is driven to migrate upward to heal the exposed layer, leaving a vacancy in the subsurface. Our transition-state calculations indicate that this surface-repair kinetic barrier is merely 0.5 eV on the clean surface (**Figure 4b**, red dashed line), a hurdle easily surmountable under ambient or operating conditions.

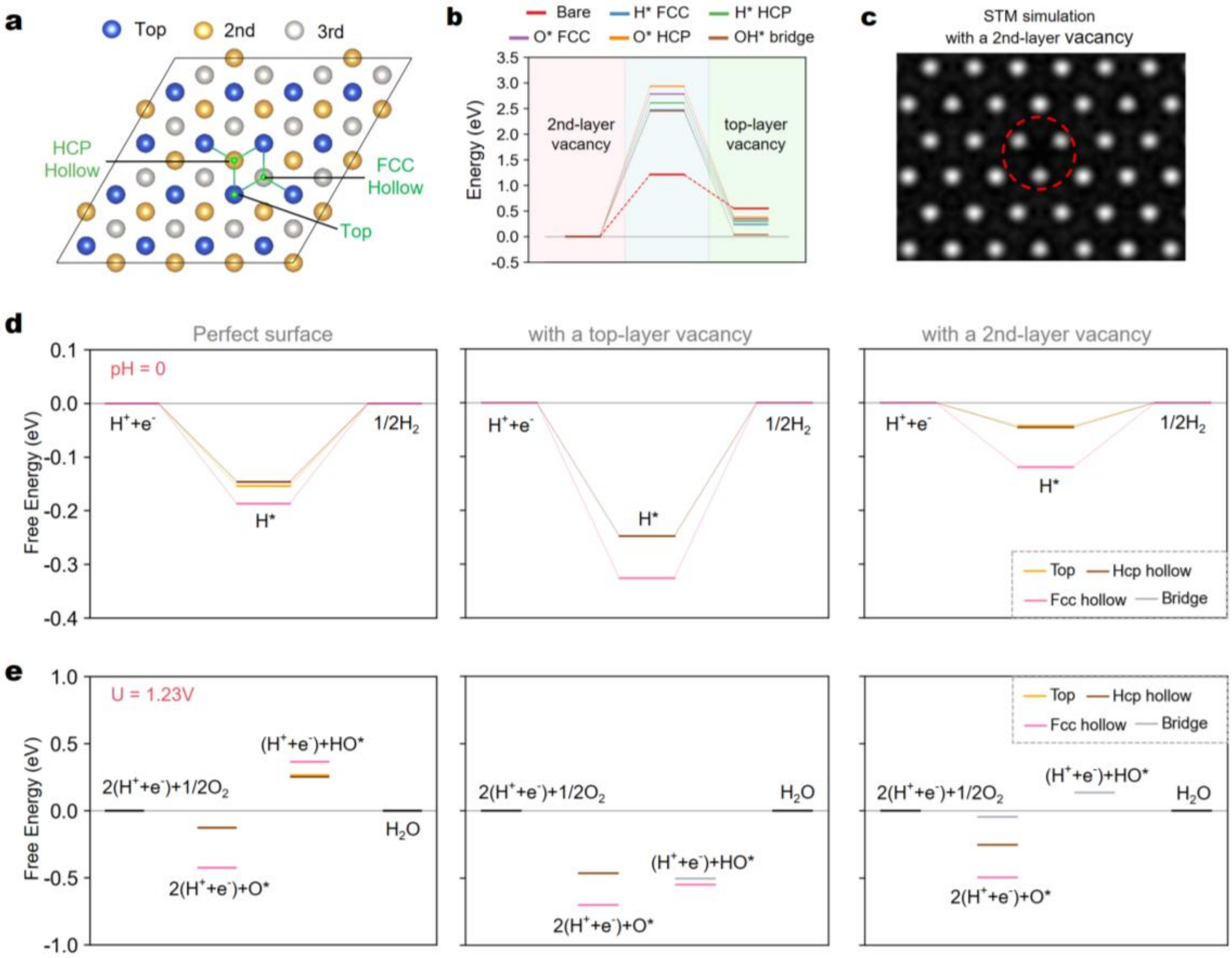

**Figure 4.** Microscopic site configurations, vacancy kinetic barriers, and electrochemical free energy landscapes for HER and ORR on Pt(111). (a) Top view of the perfect FCC (111) surface illustrating the high-symmetry adsorption sites. Lattice atoms are color-coded by depth: surface (blue), subsurface (yellow), and third layer (silver). (b) Minimum energy paths and kinetic barriers for vacancy migration from the subsurface to the topmost surface layer of Pt(111) under bare and various adsorbate-covered (H*, O*, OH*) conditions. (c) Simulated constant-current STM image of the Pt(111) surface containing a subsurface vacancy (highlighted by the red dashed circle). (d, e) Calculated Gibbs free energy diagrams for (d) the HER at pH=0, and (e) the ORR at U=1.23 V, mapped across perfect, surface-vacancy, and subsurface-vacancy surface configurations.

Upon the introduction of surface adsorbates (H*, O*, OH*), the energy difference between surface and subsurface vacancies ($\Delta E_f$) expectedly narrows. This behavior is rationalized by the enhanced chemical activity of the low-coordination surface vacancy, which binds more strongly to adsorbates than its subsurface counterpart. Crucially, however, the subsurface vacancy remains thermodynamically more stable even under full adsorbate coverage. Concurrently, the presence of these reaction intermediates drastically elevates the vacancy migration barrier from the subsurface back to the surface, raising it to well over 2 eV (**Figure 4b**). Consequently, a defective Pt surface can seamlessly repair its topmost layer by driving vacancies into the subsurface. Once buried, these defects are kinetically trapped, yielding a flat surface that is both thermodynamically and kinetically resilient under realistic reactive environments.

Importantly, this self-healing process is not merely a cosmetic surface flattening; the buried subsurface vacancies exert negligible influence on the adsorption energetics of active intermediates, thereby preserving the intrinsic catalytic performance. We mapped the calculated Gibbs free energy profiles ($\Delta E_f$) for HER and ORR across three structural configurations: the perfect surface, the surface vacancy, and the subsurface vacancy (**Figures 4**d **and 4**e). For HER (**Figure 4**d), the hydrogen intermediate (H*) exhibits near-thermoneutral adsorption energies on the perfect surface, safeguarding Pt's exceptional activity. Notably, the insertion of a subsurface vacancy alters the preferred site adsorption profile only marginally—with the baseline $\Delta G_{H*}$ shifting closely within the high-activity zone (-0.043 eV to -0.146 eV)—contrasting sharply with the severe over-binding distortion induced by an exposed surface vacancy ($\Delta G_{H*}$ < -0.3 eV).

A similarly robust trend is observed for ORR (**Figure 4**e), where both the perfect and subsurface-vacancy configurations display highly comparable adsorption energetics for oxygenated intermediates (O*, OH*). For instance, the adsorption energies of O* and OH* remain securely anchored within the optimal activity window of the ORR volcano curve, exhibiting only minor energetic deviations. In stark contrast, the surface-vacancy configuration (**Figure 4e**, middle panel) exhibits a detrimental over-binding of oxygenated species at its low-coordination defect sites. This excessive binding creates formidable free-energy barriers for subsequent proton-electron transfer steps, dropping the system far from the volcano top, which signifies severe catalytic degradation and active-site poisoning[33].

Ultimately, this self-healing mechanism allows the Pt catalyst to protect its active surface by dynamically "burying" emerging defects beneath the surface layer. This subtle thermodynamic shielding explains why subsurface vacancies on Pt(111) are notoriously difficult to resolve via experimental Scanning Tunneling Microscopy (STM), as also revealed by our simulated constant-current STM profile (**Figure 4**c).

**4. Disruption and Relief of the Reconstructed Au(100) Surface**

As established above, anomalous subsurface vacancy stabilization is an intrinsic trait of Au systems, manifesting on both the pristine and, more prominently, the reconstructed Au(111) surfaces. Experimentally, this thermodynamic inversion was originally unveiled on the reconstructed Au(100) surface, characterized by a densified, close-packed topmost layer with sixfold (quasi-hexagonal) symmetry supported on a fourfold symmetric bulk substructure[15,16]. Interestingly, literature reports indicate that this hexagonal reconstruction can be deactivated or "relieved" under high-energy electron beam or ion bombardment, reverting to a metastable, pristine (100) square lattice that requires annealing above 400 K to re-establish the reconstruction[34,35]. However, the microscopic driving force behind this defect-induced phase transition has hitherto remained elusive.

Integrating our insights on subsurface vacancy thermodynamics provides a coherent explanatory framework. High-energy particle bombardment invariably generates a high concentration of vacancies, which, according to the self-healing mechanism, are dynamically driven into the subsurface layer to maintain a flat topmost surface. However, as the accumulation of these buried subsurface defects progresses, the effective surface energy of the reconstructed phase increases monotonically with vacancy concentration (**Figure 5**). At a critical threshold, the energetic penalty of accommodating these subsurface vacancies outweighs the stabilization energy gained from surface densification.

Consequently, when the subsurface vacancy concentration reaches approximately 8%, the effective surface energy of the reconstructed phase surpasses that of the pristine, defect-free Au(100) surface (5.12 eV/nm$^2$). At this juncture, the hexagonal reconstruction undergoes a thermodynamic lifting (relief). Because these driving defects are completely buried beneath the topmost layer, this structural transition appears abruptly without obvious surface-localized precursor defects, explaining why it is exceptionally difficult to track via top-layer-sensitive imaging techniques.

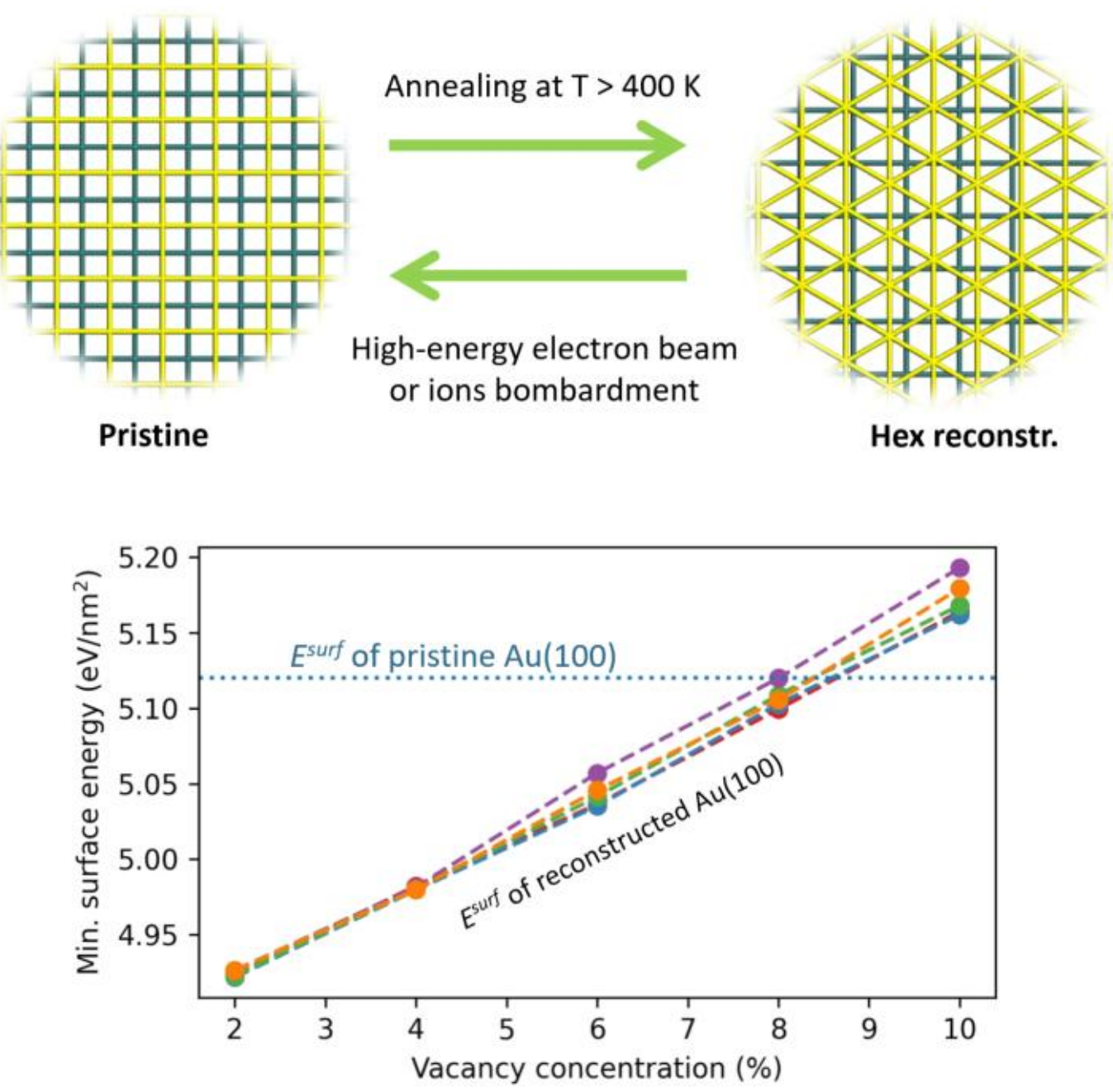

**Figure 5.** Schematic illustration of the reversible structural transition on the Au(100) surface and its underlying energetic mechanism. Upper row: Reversible structural evolution between the pristine square lattice and the densified hexagonal reconstruction under thermal annealing (T > 400 K) and high-energy particle bombardment. Lower row: Calculated minimum surface energy as a function of subsurface vacancy concentration. The dashed lines represent the evolution of the defective reconstructed surface, which crosses the baseline energy of the pristine Au(100) surface (blue dotted line) at a critical transition threshold of about 8%.

## CONCLUSION

In summary, our high-throughput screening reveals a counter-intuitive thermodynamic inversion on close-packed metal surfaces—where subsurface vacancies are preferred over surface ones—manifesting distinct physical origins in FCC (Ir, Pt, Au) and HCP (Be, Zn, Cd) metals. For the FCC trio, surface truncation triggers a profound real-space electronic reconstruction and enhanced, covalent-like intralayer bonding that penalizes surface defects. Conversely, for the HCP candidates, the anomaly is driven by their closed-shell electronic configurations interacting with local structural geometry, yielding a delocalized metallic landscape that adheres to classical empirical approximations. This thermodynamic anomaly carries profound implications under realistic conditions. On Pt(111), it underpins a dynamic "self-healing" mechanism; Gibbs free energy cascades for HER and ORR quantitatively demonstrate that burying defects underneath an intact, flat topmost layer is vital to prevent active-site poisoning and maintain optimal catalytic activity. For Au, this framework successfully decodes the critical defect threshold required to lift the long-standing Au(100) hexagonal reconstruction under radiation bombardment. Overall, our work challenges traditional coordination-based scalar models and introduces a localized electronic reconstruction paradigm essential for engineering the structural integrity of advanced catalyst surfaces.

## METHODS

**Density functional theory calculations.** All DFT calculations were systemically partitioned between the Vienna *ab initio* simulation package (VASP) version 6.4.3[36,37] and the plane-wave pseudopotential package PWmat[38,39]. To ensure the absolute intrinsic nature of the discovered anomaly, different exchange-correlation functionals were cross-examined: VASP was employed for generalized gradient approximations via the Perdew-Burke-Ernzerhof (PBE)[17], PBEsol[18], as well as the strongly constrained and appropriately norm-scaled (SCAN)[19] meta-GGA functional, both with and without the Grimme DFT-D3 empirical dispersion correction[22,23]. Conversely, the PWmat code was utilized to execute computationally demanding

Heyd-Scuseria-Ernzerhof (HSE06)[20,21] hybrid-functional calculations owing to its highly accelerated GPU-based architecture. Electron-ion interactions were modeled via the Projector Augmented Wave (PAW) method[40] in VASP and norm-conserving pseudopotentials[41] in PWmat. Given that our core metrics rely strictly on the relative formation energy differences between identical supercells, the discrepancy arising from different pseudopotential formalisms is entirely negligible. Spin polarization was explicitly enabled across all calculations to capture any potential magnetic interactions. Reciprocal space sampling was conducted using Γ-centered Monkhorst-Pack K-point grids with a dense spacing tighter than $0.02\times2\pi$ Å$^{-1}$. For VASP runs, a kinetic energy cutoff of 400 eV was enforced, with convergence thresholds set to $1.5\times10^{-2}$ eV/Å for ionic forces and $1\times10^{-8}$ eV for the electronic self-consistent field (SCF) iterations. For GPU-accelerated HSE06 calculations in PWmat, the plane-wave kinetic energy cutoff was locked at 70 Ry, with corresponding convergence metrics specified at $1.5\times10^{-2}$ eV/Å for forces and $1\times10^{-6}$ eV for electronic steps.

The pristine surface slabs consisted of five atomic layers, where the FCC (111) surface models contained 80 atoms (4×4 lateral supercell) and the HCP (0001) slabs comprised 45 atoms (3×3 lateral supercell). A vacuum gap of at least 15 Å was introduced along the perpendicular $z$-axis to prevent unphysical periodic mirror interactions. During structural optimizations, the bottom two atomic layers were frozen at their fully relaxed bulk coordinates, while the topmost three layers were allowed to dynamically relax until the maximum residual forces fell below $2\times10^{-2}$ eV/Å, under an SCF convergence threshold of $1\times10^{-5}$ eV.

Solvent effects during the HER and ORR pathways were evaluated using the implicit solvation model implemented in VASP to compute intermediate adsorption energies under realistic electrochemical environments. Scanning tunneling microscopy (STM) topographies of the subsurface defect networks were simulated via the Tersoff-Hamann approximation utilizing the VASPkit post-processing code[42] on the converged DFT charge densities.

**Transition state search.** Kinetic pathways and minimum energy paths (MEPs) for vacancy migration were identified using the Climbing-Image Nudged Elastic Band (CI-NEB) method[43]. The initial (subsurface vacancy) and final (surface vacancy) configurations were fully pre-optimized. Five discrete intermediate images were generated via linear interpolation along the reaction coordinate. To capture the stiff kinetic barriers under adsorbate coverage, the maximum force tolerance for all unconstrained atoms during CI-NEB relaxations was strictly tightened to $2\times10^{-2}$ eV/Å, with the electronic SCF criteria locked at $1\times10^{-5}$ eV to ensure precise force evaluations along the reaction coordinates.

**Crystal Orbital Hamilton Population (COHP) analysis**. To decipher the localized orbital interactions, Crystal Orbital Hamilton Population (COHP) analysis was conducted via the LOBSTER package[44]. Wavefunctions generated from VASP were projected onto localized, atom-centered Slater-type orbitals. For the late 5*d* transition metals (Ir, Pt, Au), the valence basis sets explicitly incorporated the 5*d* and 6*s* atomic orbitals to fully capture the *d*-band hybridization networks. The completeness of the projection was strictly verified, with the total quantum-mechanical absolute spilling and density spilling restricted well below 1.0% to guarantee quantitative accuracy. Bonding and anti-bonding contributions were quantified by evaluating the integrated COHP (-ICOHP) up to the Fermi level.

**EAM and MLFF calculations.** Classical empirical potential computations based on the Embedded Atom Method (EAM)[24,25] were performed via the Large-scale Atomic/Molecular Massively Parallel Simulator (LAMMPS) package[45] to serve as a baseline for isotropic metallic bonding environments. For machine learning force field (MLFF) calculations, the state-of-the-art, multi-body neural network Allegro potential was utilized, specifically adopting the pre-trained universal Allegro-OAM-L framework[46] to execute high-fidelity atom-resolved energy decompositions. For the complex reconstruction profiles of the Au(100) surface, atomistic simulations and energy evaluations were conducted using a specialized DeepMD-kit neural network potential[47] trained explicitly for capturing gold surface phase transitions[48].

**Vacancy formation energy calculation.** Following the standard thermodynamic definition, the vacancy formation energy of a specific atomic layer is expressed as[49]:

$$E_f = E_{defect} - E_{pristine} + \mu_{bulk}$$

where $E_{defect}$ and $E_{pristine}$ represent the total DFT-calculated energies of the surface slab supercell with and without an atomic vacancy, respectively, and $\mu_{bulk}$ denotes the chemical potential of the single removed metal atom, referenced to its fully relaxed bulk crystalline phase.

***d*-Band center calculation.** The $d$-band center, which characterizes the absolute and relative electronic energy states of the active layers, is mathematically defined by the first moment of the projected density of states (PDOS) as follows[50,51]:

$$\epsilon_d = \frac{\int_{E_{min}}^{E_{max}} E \cdot \rho_d(E) dE}{\int_{E_{min}}^{E_{max}} \rho_d(E) dE}$$

where $\rho_d(E)$ represents the $d$-orbital-resolved electronic density of states at a given energy level *E* relative to the Fermi energy. The integration bounds $E_{min}$ and $E_{max}$ define the energy window chosen to encompass the entire continuous $d$-band manifold.

## Data availability

The data supporting the findings of this study are available from the corresponding author upon reasonable request.

## Author contributions

YM.T. performed DFT calculations, and data analysis. P.L. conceived and supervised the project. All authors discussed the results and contributed to the manuscript.

## Acknowledgements

We acknowledge funding supports from the National Natural Science Foundation of China (Grant No. 22403104).

## Conflict of Interest

The authors declare no conflict of interest.